\documentclass[aps,prb,preprint,superscriptaddress]{revtex4}
\usepackage{latexsym} \usepackage{amsmath} \usepackage{dcolumn}
\usepackage{bm} \usepackage{graphicx} \usepackage{epsfig}
\usepackage[usenames,dvipsnames]{color}

\usepackage[]{graphicx}
\usepackage[]{wrapfig}

\begin{document}


\title{X-ray absorption near-edge spectra of overdoped
La$_{\rm 2-x}$Sr$_{\rm x}$CuO$_{\rm 4}$ high-T$_{\rm c}$ superconductors}

\author{Towfiq Ahmed} \affiliation{Department of Physics, University of
Washington, Seattle, WA 98195}

\author{Tanmoy Das} \affiliation{Department of Physics, Northeastern
University, Boston, MA 02115}

\author{J. J. Kas } \affiliation{Department of Physics, University of
Washington, Seattle, WA 98195}

\author{Hsin Lin} \affiliation{Department of Physics, Northeastern
University, Boston, MA 02115}

\author{B. Barbiellini} \affiliation{Department of Physics, Northeastern
University, Boston, MA 02115}

\author{Fernando D. Vila} \affiliation{Department of Physics, University of
Washington, Seattle, WA 98195}

\author{R.S. Markiewicz} \affiliation{Department of Physics, Northeastern
University, Boston, MA 02115}

\author{A. Bansil} \affiliation{Department of Physics, Northeastern
University, Boston, MA 02115}

\author{J. J. Rehr} \affiliation{Department of Physics, University of
Washington, Seattle, WA 98195}


\date{\today}


\begin{abstract}

We present results for realistic modeling of the x-ray absorption near edge
structure (XANES) of the overdoped high-$T_{\rm c}$ superconductor La$_{\rm
2-x}$Sr$_{\rm x}$CuO$_{\rm 4}$ in the hole doping range x =
0.20-0.30. Our computations are based on a real-space Green's
function approach in which strong-correlation effects are taken
into account in terms of a doping-dependent self-energy. The
predicted O K-edge XANES is found to be in good accord with
the corresponding experimental results in this overdoped regime.
We find that the low energy spectra are dominated by the contribution
of O-atoms in the cuprate planes, with little contribution from apical
O-atoms.

\end{abstract}

\maketitle

\section{INTRODUCTION}

In their undoped parent compounds, high-$T_c$ cuprate superconductors are
antiferromagnetic insulators which are characterized by a gap driven by
strong electron correlations. For this reason these materials are
commonly referred to as Mott insulators. Strong correlation effects weaken
with increased electron or hole doping, eventually yielding a metallic state.
In La$_{2-{\rm x}}$Sr$_{\rm x}$CuO$_{\rm 4}$ (LSCO), for example, at a hole doping
level of x $\sim$ 0.16, a paramagnetic state emerges and the material appears
to recover Fermi liquid properties.
However, despite over two decades of
intense experimental
and theoretical effort, the underlying principles governing how a
Mott insulator transitions into a Fermi-liquid with doping are still not
well understood.\cite{tanmoy} The answer seems to be hidden within
the mechanisms through which the quasiparticle spectral weight passes from
the insulating Mott-Hubbard bands to the in-gap states near the Fermi
level. In electron doped cuprates, the Mott gap and the associated lower
Hubbard band can be directly probed by photoemission spectroscopy.\cite{Kusko,ASWT}
In the hole doped cuprates, on the other hand, this gap lies above the
Fermi energy, so that techniques sensitive to empty states within a few
eV above the Fermi energy must be deployed. Accordingly, light scattering
techniques have been used, including optical, resonant inelastic
x-ray scattering (RIXS), and x-ray absorption near edge spectroscopy
(XANES) which probes the density of states (DOS) of empty states above
the Fermi energy via excitations from core levels.\cite{xanes,bobrixs,bobZrixs,tanmoy}

The purpose of this study is to model the XANES spectrum of
LSCO realistically, as an exemplar hole-doped cuprate, and to compare
and contrast our
theoretical predictions with available experimental data. The analysis
is carried out using a real-space Green's function (RSGF) approach as
implemented in the FEFF9 code.\cite{rehrpccp,rehr2009} Strong correlation
effects on the electronic states near the Fermi energy ($E_{\rm F}$) are
incorporated by adding additional self-energy corrections to the one-particle
electron and hole propagators.  We concentrate in this initial study on the
overdoped system because the cuprates are in a paramagnetic state in this
doping range. Consequently the treatment of correlations effects is simpler,
due to the absence of the pseudogap in the electronic spectrum. In LSCO, the 
pseudogap is found to vanish near x$=0.20$.\cite{tanmoy,ASWT} We start with a
generic plasmon-pole self-energy and then dress our calculations with a doping
dependent paramagnetic self-energy $\Sigma$ obtained within the
self-consistent quasiparticle-GW (QP-GW)
scheme.\cite{bobparamagnon,tanmoy,bobCorpes}
Here GW refers to the Hedin approximation to the self energy
$\Sigma=iGW$, where $G$ is the one-particle
Green's function and $W$ the screened Coulomb interaction.\cite{HedinGW}
This self-energy has been shown
to capture key features of strong electronic correlations in various
cuprate spectroscopies including ARPES \cite{susmita}, RIXS
\cite{susmitarixs}, optical\cite{tanmoy} and neutron
scattering\cite{xanes}, in good agreement with experiments in both
electron and hole doped systems.

Many key properties of cuprates, including the physics of
superconductivity, involve hybridized Cu $d_{x^2-y^2}$ and O $p_{x,y,z}$
orbitals near the Fermi energy $E_F$.
Thus the natural choices for the probe
atoms in which the incoming x-ray excites a core-hole are Cu and O. Since
dipole selection rules do not allow K-edge excitations in Cu atoms to
couple to $d$-bands, we focus here mainly on the O K-edge XANES.
This edge may be expected to reflect doping dependent changes in the near $E_F$
spectrum  through its sensitivity to the O-$p$ states. In O K-edge XANES
experiments on LSCO,\cite{fink_5,fink_3,fink_4} two `pre-peaks' have been 
observed
to vary with Sr concentration. Our analysis indicates that the energy
separation between these two peaks, which is comparable to the optical gap
in the insulating phase,\cite{tanmoy} is associated with the Mott gap.\cite{Mottgap}
In particular, the upper XANES peak corresponds to the empty states of the
upper Hubbard band, and the lower peak to empty states in the lower
Hubbard band resulting from hole doping. With increasing doping, the
lower peak, which is absent for x = 0., starts to grow while the upper peak
loses intensity.  In the overdoped regime, the lower peak reaches a
plateau,\cite{peets_1} while the intensity of the upper peak is
substantially suppressed.


In order to assess effects of core-hole screening, we have also calculated
the Cu K-edge XANES, again using the same RSGF
approach.\cite{rehrpccp,rehr2009} In this connection, two different core-hole
models were considered: (i) Full screening, i.e., without a core-hole
as in the ``initial state rule" (ISR),\cite{isr2004} and (ii) RPA screening as is typically
used in Bethe-Salpeter equation (BSE) calculations.\cite{rpa2005} 
Both of these
core-hole models reasonably reproduce the experimental XANES of the O
K-edge in the pre-peak region in overdoped LSCO. 

The remainder of this article is organized as follows. Introductory remarks
in Section I are followed by a brief account of the methodological details
of the RSGF formalism in Section IIA, and of the QP-GW self-energy
computations in Section IIB. XANES results based on the plasmon-pole
self-energy are discussed in Section IIIA, while results based on doping
dependent QP-GW self-energies are taken up in Section IIIB. Finally Section
IV contains a summary and conclusions.

\section{Theory}

\subsection{Real-space Green's function formulation}

Here we briefly outline the real-space Green's function 
multiple-scattering formalism underlying the FEFF code. More detailed
accounts are given elsewhere.\cite{rehrpccp,rehr2009} The quasi-particle
Green's function for the excited electron at energy $E$ is defined as
\begin{equation}
G(E)=\left[{E - H
-\Sigma(E)}\right]^{-1}.
\end{equation}
Here $H$ is the independent particle (i.e., Kohn-Sham) Hamiltonian, 
\begin{equation}
H={\frac{p^{2}}{2}}+V,
\end{equation}
with $V$ being the Hartree potential plus a ground state exchange-correlation
density functional, which in FEFF9 is taken to be the von Barth-Hedin
functional.\cite{Barth-Hedin}  Throughout this paper Hartree atomic units
($e=\hbar=m=1$) are implicit. 
This Hamiltonian together with the Fermi-energy $E_F$ are calculated
self-consistently using the RSGF approach outlined below.
In Eq.\ (1) the quantity $\Sigma(E)$ is the energy-dependent one-electron 
self-energy. In this work we use a GW self-energy designed to 
incorporate the strong-coupling effects in cuprates, as discussed further 
in Section IIB below.  

In the RSGF approach it is useful to decompose the total Green's function
${\it {G}}(E)$ as
\begin{equation}
G(E) = G^{c}(E)
+ G^{sc}(E),
\end{equation}
where ${G^{c}(E)}$ is the contribution from the central
atom where the x-ray is absorbed and ${G^{sc}(E)}$ is the scattering part.
For points within a sphere surrounding the absorbing atom
the angular dependence of the real space Green's function
can be expanded in spherical harmonics as
\begin{equation}
G({\bf {r}},{\bf {r'}},E) = \sum_{L,L'} Y_L(\hat{\bf r}) \, G_{L,L'}(r,r',E) \, Y^{*}_{L'}(\hat{\bf r'}).
\end{equation}
Here, $Y_L$ is a spherical harmonic and $L=(l,m)$
denotes both orbital and azimuthal quantum numbers.
The physical quantity measured in XANES for x-ray photons of polarization
${\bf \hat \epsilon}$ and energy $\omega=E-E_c$ is the x-ray absorption
coefficient $\mu(\omega)$, where $E_c$ is the core electron energy,
and $E$ is the energy of the excited electron. The absolute edge
energy is given by $\omega = E_F-E_c$, where $E_F$ is the Fermi level.
The FEFF code calculates
both $\mu(\omega)$, the site- and {\it {l}}-projected DOS
$\rho^{(n)}_{l}(E)$ at site $n$
 and the Fermi energy ($E_F$) self-consistently.  
These quantities can be expressed in terms of the Green's
function in Eq.\ (4) as
\begin{equation}
\mu(\omega) \propto \ - \frac{2}{\pi} {\rm Im}\, \left< \phi_0|{\hat {\bf
{\epsilon}}}.{\bf {r}} G({\bf{r}} , {\bf{r'}} , E) {\hat
{\bf{\epsilon}}}.{\bf {r'}}|\phi_0 \right> ,
\end{equation}
where $E=\omega+E_c$ and
\begin{equation}
\rho_l^{(n)}(E) = \ - \frac{2}{\pi} {\rm Im}\, \sum_m \int_{0}^{R_n} G_{L,L}(
r,r, E) \, r^2 \, dr \ ,
\end{equation}
respectively. Here $\left|\phi_0\right>$ is the initial state of the absorbing atom and
$R_n$ is the Norman radius\cite{feff84ref} around the n$^{th}$ atom, which is analogous to the
Wigner-Seitz radius of neutral spheres, and the factor 2 accounts for spin
degeneracy. 


%

\subsection{Self-energy corrections from strong correlations in
cuprates}

In the optimal or overdoped regime of present interest, cuprates do not
exhibit any signature of a symmetry-breaking order parameter, and thus the
quasiparticle dispersion can be well-described with a paramagnon-renormalized
one band Hubbard model.\cite{tanmoy} The details of the
related self-energy are given in Ref.~\onlinecite{bobparamagnon} and are
summarized here for completeness.

In our QP-GW approach, we start with a tight-binding single band bare
dispersion $\xi_{\bf k}$ where the parameters are obtained by fitting to
the first-principles LDA dispersion.\cite{Arun3,foot2,Zeit} The values of the tight-binding
parameters used in LSCO are from Ref.~\onlinecite{tanmoy}.  The
LDA dispersion is then self-consistently dressed by the full spectrum of
spin and charge fluctuations treated at the RPA level. The susceptibility
can be written in terms of the bare susceptibility $\chi_0({\bf q},\omega)$
as
\begin{equation}
\chi({\bf q},\omega) = \frac{\chi_0({\bf q},\omega)}
{1-{\bar U}\chi_0({\bf q},\omega)}.
\end{equation}
Here ${\bar U}$ is the renormalized Hubbard $U$ value.
The imaginary part of
the RPA susceptibility provides the dominant fluctuation interaction
to the electronic system, which can be represented by $W({\bf
q},\omega)=(3/2){\bar U}^2\chi^{\prime\prime}({\bf q},\omega)$. The
resulting self-energy correction to the LDA dispersion within the GW
approximation is
\begin{eqnarray}
\Sigma ({\bf k},\omega) &=& Z\sum_{\bf q}W({\bf q},\omega')\Gamma({\bf k},{\bf q},\omega,\omega')\nonumber\\
&\times&\left[ \frac{f({\bar \xi}_{{\bf k}-{\bf q}})}
{\omega+\omega'+i\delta-{\bar \xi}_{{\bf k}-{\bf q}}} +\frac{1-f({\bar \xi}_{{\bf k}-{\bf q}})}
{\omega-\omega'+i\delta-{\bar \xi}_{{\bf k}-{\bf q}}}\right],
\end{eqnarray}
where $f(\xi)$ is the Fermi function and $\Gamma$ is the vertex correction
defined below.

Different levels of self-consistency within the GW scheme involve
different choices for $\chi_0$ and dispersion ${\bar \xi}_{\bf k}$.
Within our QP-GW
scheme, the Green's function entering into the $\chi_0$ bubble is
renormalized by an approximate renormalization factor $Z$ which is evaluated self-consistently.
The corresponding vertex correction is taken within the Ward identity
$\Gamma=1/Z$, and ${\bar
\xi}_{\bf k}=Z(\xi_{\bf k}-\mu)$ is the renormalized dispersion where
$\mu$ is the chemical potential. The renormalized band is employed to
calculate the full spectrum of spin susceptibility:
\begin{equation}
\chi_0({\bf q},\omega) = -Z\sum_{\bf k} \frac{f({\bar \xi}_{\bf k})-f({\bar \xi}_{{\bf k}+{\bf q}})}
{\omega+i\delta+{\bar \xi}_{\bf k}-{\bar \xi}_{{\bf k}+{\bf q}}}.
\end{equation}

The doping dependence of $U$ is discussed in
Refs.~\onlinecite{tanmoy} and \onlinecite{xanes}. A single,
universal function
${\bar U}(x)$ is found to reasonably describe a number of
different spectroscopies including
photoemission, optical spectra\cite{tanmoy} and the present XANES. This doping
dependence is consistent with charge screening.\cite{bobrixs}

\section{Results and Discussion}
\begin{figure}
\includegraphics[scale=0.34,angle=-90]{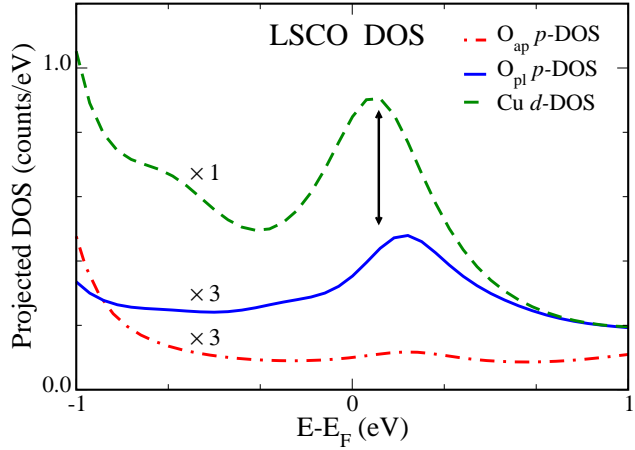}
\caption[example]
{(color online)
Site- and $\it l$-projected DOS in doped LSCO, x$=0.30$.
Cu-$d$ (green dashed line),
O$_{pl}$-$p$ (blue solid line) and O$_{ap}$-$p$ (red dashed-dotted line) DOS are shown. Note scaling of
the lower two curves by a factor of 3. 
Arrows mark the structure related to 
the van Hove singularity in the hybridized Cu-O bands discussed in the
text. 
}
\end{figure}

In Subsection IIIA below, we discuss O and Cu K-edge XANES using a
generic GW plasmon-pole self-energy and RPA core-hole screening, but without
the self-energy correction arising from strong correlation effects. Subsection
IIIB examines doping-dependent effects of self-energy corrections
on the O K-edge XANES.

\subsection{XANES without self-energy corrections}

\begin{figure}
\includegraphics[scale=0.34,angle=-90]{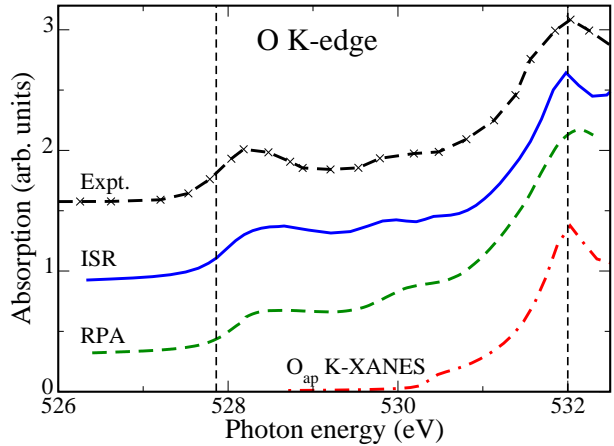}
\caption
 {(color online)
Theoretical and experimental\cite{fink_2} O K-edge XANES spectra in
overdoped LSCO ($x$=0.30). Computations where the core-hole is fully
screened, i.e. using the initial state rule (ISR) (blue solid line) are compared with
results based on RPA screened core-hole (green dashed line) and experiments
(dashed crossed line). The vertical dashed line at 527.8 eV marks the
edge energy $E_F-E_c$ which is sensitive to doping.
A vertical dashed line is also drawn through the feature at 532 eV.
 This feature is due to
the apical O, as shown by the red dashed-dotted line, calculated with the ISR.  
}
\end{figure}
\begin{figure}
\includegraphics[scale=0.34,angle=-90]{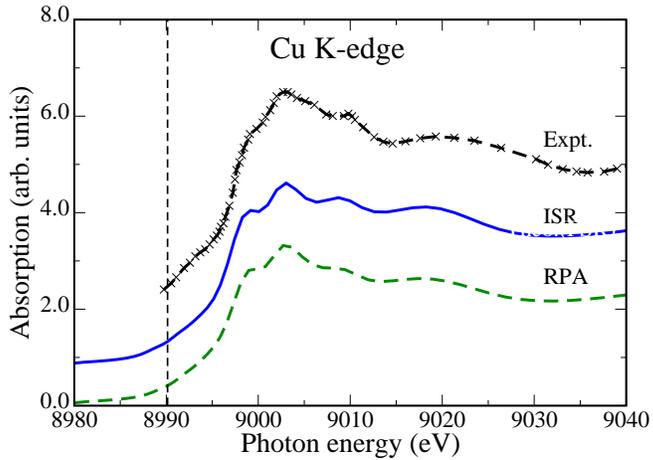}
 \caption
 {(color online)
Theoretical and experimental\cite{kosugi_1} Cu K-edge XANES spectra in
undoped LSCO ($x$=0.).  Curves have same meaning as in Fig.\ 2.
The vertical dashed line at 8990 eV marks the edge energy $E_F-E_c$ 
as in Fig.\ 2.
}
\end{figure}

Since K-edge XANES probes the site-dependent $p$-density of states ($p$-DOS),
we consider first the projected $p$-DOS from O and Cu sites near $E_F$
as obtained using the FEFF code. The low-temperature orthorhombic crystal
structure with space group {\rm Bmab} was used.\cite{lsco_structure}
It is important to note that the structure
involves two inequivalent O-atoms with different chemical environments,
namely, the O-atoms in the cuprate planes (O$_{pl}$), and the apical O-atoms
(O$_{ap}$) which lie in the La-O planes.\cite{foot3,BCPA} The projected densities of states on
the two distinct O-atoms and the Cu-atoms are shown in Fig.\ 1. Also shown
for reference is the projected $d$-DOS from Cu-sites, even
though as already pointed out K-edge excitations do not couple with the
$d$ angular momentum channel. The $p$-DOS from the O-atoms in the cuprate
planes (middle curve) is seen to display the structure marked by arrows,
arising from the hybridization of localized Cu {{\it d}} and O {{\it p}}
orbitals, which is related to the van Hove singularity (VHS) near $E_F$ in the
band structure of LSCO. The O$_{\rm ap}$ contribution is seen to be quite
small and structureless within an energy range $\pm 0.75$ eV of {\it
{E$_{\rm F}$}}, although a weak replica of the VHS peak suggests a small hybridization of the apical and in-plane oxygens. This implies that O K-edge XANES pre-peak is mainly
associated with unoccupied electronic states from atoms lying in the Cu-O
planes. Notably, the Cu-$p$ DOS (not shown) in the near $E_F$ energy window of Fig.\ 1 is
also quite small and structureless and becomes significant only several
eV above $E_F$. The experimental evidence for the aforementioned
features of XANES spectra has been discussed previously by several authors.
\cite{fink_3,fink_4,fink_5}

The treatment of core-hole screening in the O K-edge spectrum is addressed in
Fig.\ 2, where we compare the experimental results from overdoped LSCO
with computations using two different core-hole
screening models. The computed XANES spectrum using a fully screened hole (blue
solid curve) is seen to be in good accord with the results of the RPA screened
hole (green dashed curve), although the intensity of the feature at 532 eV
differs somewhat in the ISR and RPA results. All three curves in Fig.\ 2
show the presence of the pre-edge peak around 528.5 eV, a weaker
pre-edge feature around 530 eV, and the prominent peak at 532 eV, which is due to the apical oxygen, as demonstrated by the calculated partial absorption, red (dashed-dotted) curve.  
Fig.\ 3 presents results along the preceding lines for the Cu K-edge XANES in
undoped LSCO. Since the Cu K-edge probes $p$-states which are only weakly
correlated, it provides a useful check of our theoretical method for the case of
weak-correlation.
The results in Figs.\ 2 and 3 clearly indicate that
the K-edge spectra are not sensitive to the core-hole screening model, as both
the RPA and the ISR give results in reasonable agreement with experiment.

\subsection{Strong correlation effects and doping dependence}

\begin{figure}
\includegraphics[scale=0.34,angle=-90]{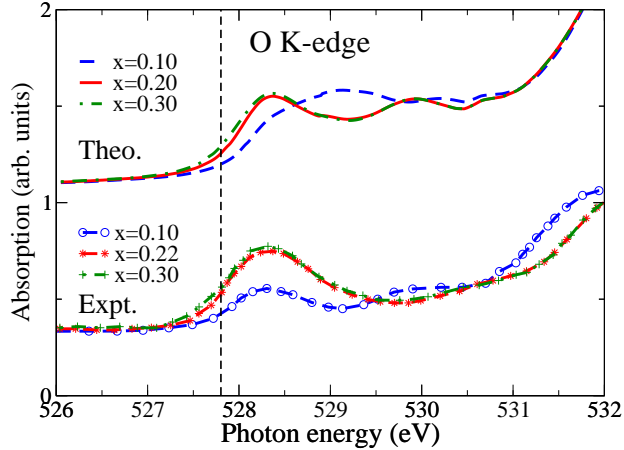}
 \caption[example]
  {(color online)
Doping-dependent theoretical (upper curves) and
experimental\cite{peets_1} (lower curves) O K-edge XANES spectra in
LSCO. Computations include the effect of self-energy corrections of Eq.\ (8)
resulting from strong correlation effects. Different dopings with different 
concentrations of x are
shown by lines of various colors (see legend). The vertical line marks the
edge energy $E_F-E_c$ as in Fig.\ 2. 
}
\end{figure}
\begin{figure}
\includegraphics[scale=0.34,angle=-90]{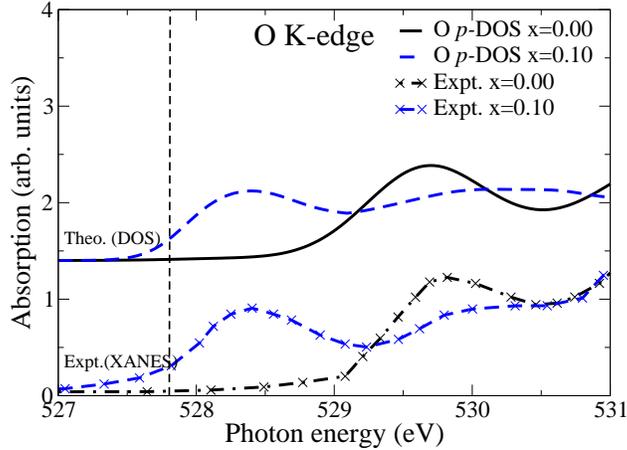}
\caption
 {(color online)
O K-edge XANES from 
experiment\cite{chen_1} (lower set of broken curves) for x = 0. and 
x = 0.10, compared with theoretical single-band $p$-DOS
(upper curves) for O$_{pl}$ for x = 0. and x = 0.10.\cite{xanes}
The vertical line marks the edge energy $E_F-Ec$ as in Fig.\ 4.
}
\end{figure}

We emphasize that the conventional LDA-based formalism is fundamentally
limited in its ability to describe the full doping dependence of the
electronic structure of the cuprates, because the LDA yields a metallic
instead of an insulating state in the undoped system. Therefore, strong
correlation effects must be added to properly model the doping evolution
of electronic states. Fig.\ 4 compares recent experimental
results\cite{peets_1} with the theoretical O K-edge XANES spectra in
overdoped LSCO, where the self-energy correction $\Sigma$ along the lines
of Section IIB is included in the computations. Representative real and
imaginary parts of $\Sigma$ are shown in Ref.~\onlinecite{tanmoy}. 
Focusing first on the overdoped regime for hole doping concentration between
x = 0.20 to x = 0.30, both theory and experiment display a small, systematic
shift of the edge to lower energies with increasing doping. 
The low energy peak at 528.5 eV in 
Fig.\ 4 shows the experimental and theoretical edge regimes,
showing that both display a similar shift of the Fermi level with doping.
Otherwise, the XANES spectra undergo relatively little change in the
overdoped system.

As the doping is further reduced, the intensity of the 528.5 eV peak
decreases while a new peak appears near 530 eV and rapidly grows with
underdoping until at x = 0., the 528.5 eV peak is completely gone.  The
remaining 530 eV peak represents the upper Hubbard band, and its shift in
energy from the Fermi level is consistent with optical 
measurements.\cite{tanmoy}
Turning to the x = 0.10 spectra in Fig.\ 4, we see that, as expected, now
theory differs substantially from experimental results. Although theory
correctly reproduces the reduced intensity of the 528.5 eV peak, it does
not show the observed enhanced intensity of the upper peak at 530 eV.
Instead, the spectral weight is shifted halfway between the lower and
upper peaks. In Fig.\ 5, the experimental results indicate the opening 
of a gap in the
spectrum which is
not captured in our modeling. However, we were able to reproduce the
experimental doping dependence\cite{peets_1} in a simpler calculation in
which the XANES spectrum is modeled via the empty density-of-states, but
in which self-energy corrections including the magnetic gap are
incorporated.\cite{xanes} Fig.\ 5 compares the experimental XANES
spectrum with this DOS approximation at x = 0.10.  The splitting of the
spectrum into two peaks with the appropriate gap is well reproduced.  It
should be noted that this same self energy reproduced the optical and RIXS
gaps as a function of doping.\cite{tanmoy,bobrixs,susmitarixs}

Interestingly, the good agreement between theory and experiment in the
overdoped case implies that the upper peak at 530 eV possesses little
weight in the overdoped system. In particular, we find a saturation 
of the lower energy feature (~528eV) in the O K-edge spectra [Fig.\ 4] as the doping value is increased
 beyond the optimal doping, except for the Fermi edge shift.  This saturation
 is in good agreement with the experiment by Peets
et al.,\cite{peets_1} but is inconsistent with model calculations by 
Liebsch\cite{Liebsch} and
Wang et al.\cite{Wang} This effect requires a transfer of
spectral weight from the upper to the lower peak, which is much larger
than that predicted by {\it t-J} models of the cuprates.\cite{xanes}
A similar conclusion
was arrived at by Peets {\it et al.}\cite{peets_1} Such anomalous
spectral weight transfer has been discussed in several spectroscopies of
cuprates.\cite{tanmoy,sawatzky_1,comanac}  The present intermediate coupling
calculations can provide a satisfactory explanation of these
results.\cite{ASWT}

\section{Summary and Conclusions}

We have carried out a realistic calculation of the XANES of the
cuprate superconductor LSCO. Our focus in this first such attempt is on the O
K-edge XANES in overdoped LSCO over the hole doping range x = 0.20 to
x = 0.30. For this purpose, we have incorporated doping dependent
self-energy corrections due to strong electron correlations into 
a real space Green's function formalism, which is implemented in
the FEFF9 code. Good agreement is
found between theoretical and experimental doping evolution of the XANES
in overdoped LSCO. Although LSCO has two chemically distinct
O-atoms, the O K-edge XANES is dominated by the contribution of O-atoms in
the cuprate planes. Apical O-atoms only begin to make a significant contribution
to the spectrum at higher energies greater than 531~eV.
In examining effects of the
core-hole screening, we find that the spectra are insensitive to the 
core-hole screening model
at least in the overdoped regime. In the
underdoped case, as expected, our self-energy corrections, which are
appropriate for the overdoped paramagnetic system, fail to correctly
describe salient features of the spectra. A simple calculation suggests that 
correcting this will require a more comprehensive modeling of XANES including
effects of pseudogap physics on the self-energies in the underdoped regime.

\section{ACKNOWLEDGMENT}

This work is supported by the Division of Materials Science \&
Engineering, Basic Energy Sciences, US Department of Energy
Grants DE-FG03-97ER45623 and DE-FG02-07ER46352, and has benefited from
the allocation of supercomputer time at NERSC and
Northeastern University's Advanced
Scientific Computation Center (ASCC). The research also benefited from the
collaboration supported by the Computational Materials Science Network
(CMSN) program of US DOE under grant DE-FG02-08ER46540.


\begin{thebibliography}{24}
\bibitem{Mottgap}We use the term `Mott gap' to denote the pseudogap in the CuO$_2$ plane.
\bibitem{foot2}Effects of doping are treated at the rigid-band level.\cite{Zeit}
\bibitem{foot3}It will be interesting to examine effects of La/Sr substitution on apical O sites beyond the rigid band model using first-principles approaches.\cite{BCPA}
\end{thebibliography}

\end{document}